\documentclass[letter]{aa}

\usepackage{graphicx}
\usepackage{txfonts}
\usepackage[]{hyperref}
\begin{document}

\newcommand{\lya}{Lyman-$\alpha$}

\title{The high-energy environment and atmospheric escape of the mini-Neptune K2-18~b\thanks{The \emph{HST} Lyman-$\alpha$ spectra are available in electronic form at the CDS via anonymous ftp to cdsarc.u-strasbg.fr (130.79.128.5) or via http://cdsweb.u-strasbg.fr/cgi-bin/qcat?J/A+A/}}

\author{Leonardo A. dos Santos
\inst{1}
\and
David Ehrenreich\inst{1}
\and
Vincent Bourrier\inst{1}
\and
Nicola Astudillo-Defru\inst{2}
\and
Xavier Bonfils\inst{3}
\and
Fran\c{c}ois Forget\inst{4}
\and
Christophe Lovis\inst{1}
\and
Francesco Pepe\inst{1}
\and
St\'ephane Udry\inst{1}
}

\institute{Observatoire astronomique de l’Université de Genève, 51 
chemin des Maillettes, 1290 Versoix, Switzerland\\
\email{Leonardo.dosSantos@unige.ch}
\and
Departamento de Matem\'atica y F\'isica Aplicadas, Universidad Cat\'olica de la Sant\'isima Concepci\'on, Alonso de Rivera, 2850 Concepci\'on, Chile
\and
Universit\'e Grenoble Alpes, CNRS, IPAG, 38000 Grenoble, France
\and
Laboratoire de M\'et\'eorologie Dynamique, Institut Pierre Simon Laplace, Universit\'e Paris 6 Boite Postale 99, 75252 Paris cedex 05, France
}

\date{Received 6 December 2019; accepted 13 January 2020}

 
\abstract{K2-18~b is a transiting mini-Neptune that orbits a nearby (38 pc), cool M3 dwarf and is located inside its region of temperate irradiation. We report on the search for hydrogen escape from the atmosphere K2-18~b using \lya\ transit spectroscopy with the Space Telescope Imaging Spectrograph (STIS) instrument installed on the \emph{Hubble Space Telescope} (\emph{HST}). We analyzed the time-series of fluxes of the stellar \lya\ emission of K2-18 in both its blue- and redshifted wings. We found that the average blueshifted emission of K2-18 decreases by $67\% \pm 18\%$ during the transit of the planet compared to the pre-transit emission, tentatively indicating the presence of H atoms escaping vigorously and being blown away by radiation pressure. This interpretation is not definitive because it relies on one partial transit. Based on the reconstructed \lya\ emission of K2-18, we estimate an EUV irradiation in the range $10^1-10^2$ erg s$^{-1}$ cm$^{-2}$ and a total escape rate on the order of $10^8$ g s$^{-1}$. The inferred escape rate suggests that the planet will lose only a small fraction ($< 1$\%) of its mass and retain its volatile-rich atmosphere during its lifetime. More observations are needed to rule out stellar variability effects, confirm the in-transit absorption, and better assess the atmospheric escape and high-energy environment of K2-18~b.}


\maketitle
%

\section{Introduction}

Short-period exoplanets orbiting nearby, cool M dwarfs are prime targets for the search and characterization of atmospheres of low-mass, sub-Neptune-sized worlds. One particular target that falls in this category is K2-18~b, which was first pointed out as a transiting planet candidate by \citet{2015ApJ...809...25M} and later confirmed with \emph{Spitzer} photometry \citep{2017ApJ...834..187B} and Doppler velocity measurements \citep{2017A&A...608A..35C}. This planet has a radius of $R_\mathrm{P} = 2.711 \pm 0.065$ R$_\oplus$, a mass of $M_\mathrm{P} = 8.64 \pm 1.35$ M$_\oplus$ and an orbital period of $T_\mathrm{orb} = 32.9$ d \citep{2019A&A...621A..49C}. At an average distance of 0.14 au from its host star, K2-18~b receives a similar amount of bolometric irradiation to that received by the Earth from the Sun. Nevertheless, its high-energy environment is unconstrained, and its density is consistent with either a significant H$_2$/He envelope or a $100\%$ H$_2$O composition \citep{2018AJ....155..257S, 2019A&A...621A..49C}. The host star is a nearby M2.8-type dwarf located at 38 pc \citep{2018A&A...616A...1G}, rendering K2-18~b one of the best mini-Neptunes suitable for atmospheric follow-up using the \emph{Hubble Space Telescope} (\emph{HST}), the \emph{James Webb Space Telescope} (\emph{JWST}), and high-resolution infrared spectrographs.

The 0.4--5.0 $\mu$m transmission spectrum of K2-18~b measured with data from the \emph{Kepler} satellite (\emph{K2} mission), the Wide-Field Camera 3 (WFC3/\emph{HST}), and the Infrared Array Camera (IRAC/\emph{Spitzer}) revealed the presence of water vapor in its lower atmosphere \citep{2019ApJ...887L..14B, 2019NatAs.tmp..451T}. By comparing atmospheric models with the data, \citeauthor{2019ApJ...887L..14B} concludes that the best match is a H$_2$-dominated atmosphere with water vapor absorbing above the cloud deck below the 10-1000 mbar pressure level. While these observations provided us with some initial information regarding the composition of its lower atmosphere, they do not constrain the abundances of molecular species.

Models predict that the deposition of high-energy photons (X-rays and far-ultraviolet) produced by the host star leads to an expansion of the planetary upper atmosphere, as well as the production of H atoms due to photodissociation of H$_2$O \citep[e.g.,][]{1983ApJ...264..726I, 1993JGR....98.7415W}. This expansion populates the outer layers of the planetary atmosphere where the gas is collisionless, also known as the exosphere. It is therefore likely that the atmosphere of K2-18~b, which is rich in H$_2$ and H$_2$O, possesses a H-rich exosphere. Previous \emph{HST} observations have shown evidence for the presence of large-scale, H-rich exospheres around the warm Neptunes GJ~436~b \citep{2015Natur.522..459E, 2017A&A...605L...7L, 2019A&A...629A..47D} and GJ~3470~b \citep{2018A&A...620A.147B}. However, to date, no evidence for extended atmospheres has been found for planets smaller than Neptune, as non-detections were reported for the super-Earths 55~Cnc~e \citep{2012A&A...547A..18E}, HD~97658~b \citep{2017A&A...597A..26B}, GJ~1132~b \citep{2019AJ....158...50W}, and $\pi$ Men c \citep{2019arXiv191206913G}, and marginal detections were reported for the small rocky-planet systems in TRAPPIST-1 \citep{2017AJ....154..121B} and Kepler-444 \citep{2017A&A...602A.106B}.

In this letter, we report the results of a series of far-ultraviolet (FUV) observations of two transits of K2-18~b using the Space Telescope Imaging Spectrograph (STIS) installed on \emph{HST}. The aim of these observations was to perform Lyman-$\alpha$ transmission spectroscopy of K2-18~b in order to detect its H-rich exosphere, which should produce an excess absorption in the stellar Lyman-$\alpha$ emission during the transit of the planet.

\section{Observations and data reduction}

\begin{figure*}
\centering
\begin{tabular}{cc}
\includegraphics[width=0.44\hsize]{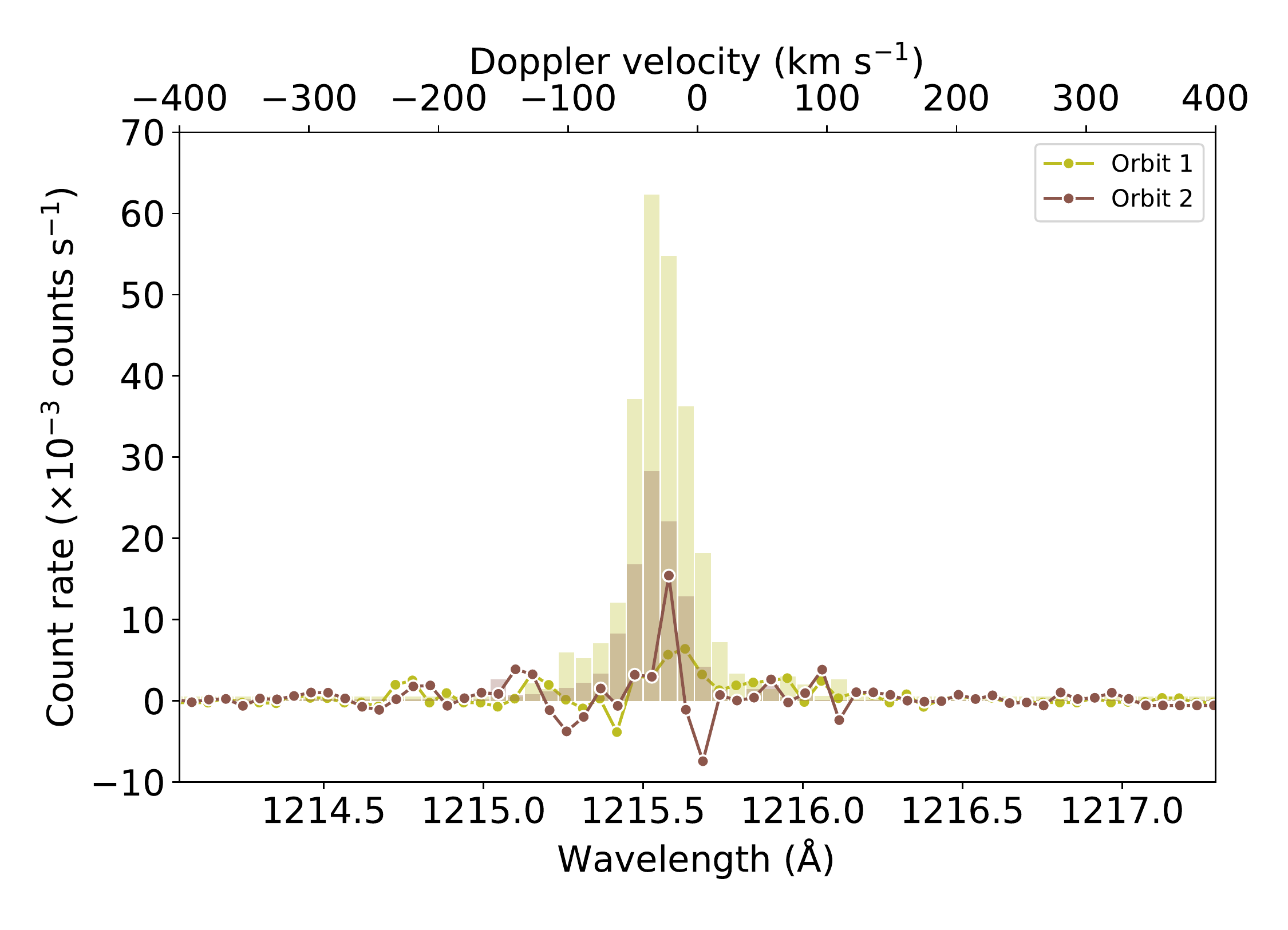} & \includegraphics[width=0.44\hsize]{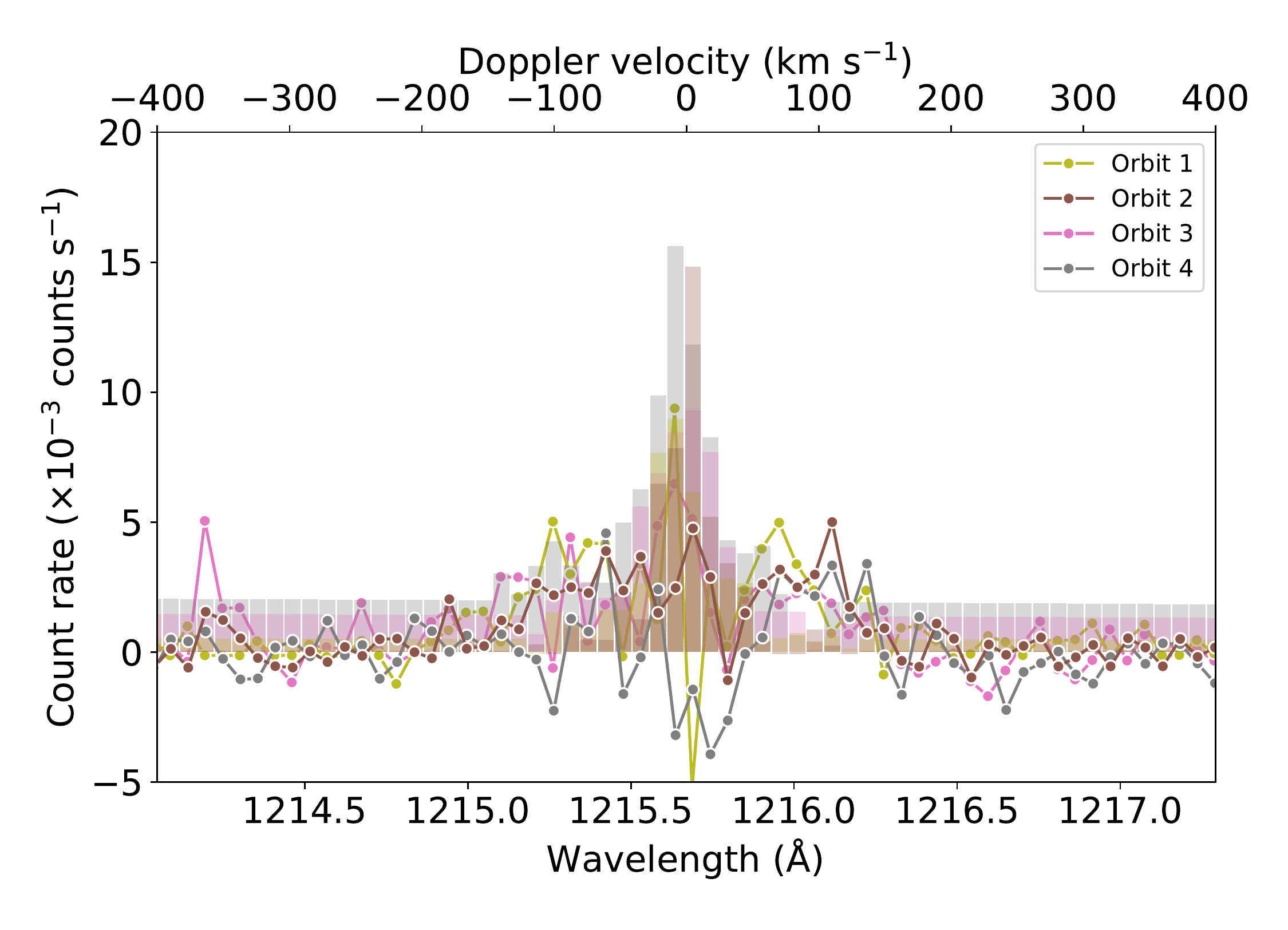}
\end{tabular}
\caption{\emph{HST}/STIS spectra of K2-18 in Visits A (left panel) and B (right panel) obtained after data reduction (lines). The geocoronal and background contamination is shown as vertical bars; in Visit A, the geocoronal emission overwhelms the stellar fluxes, preventing us from reliably measuring the latter. The Doppler velocities are in the stellar rest frame.}
\label{k2-18_A}
\end{figure*}

\object{K2-18} was observed with \emph{HST}/STIS and the grating G140M (resolving power $R \approx 10000$) during two transits on June 18 2017 and March 9 2018 (Program GO-14221, PI: D. Ehrenreich). The first visit (A) allowed five exposures, of which only the first two were successful but were severely contaminated by strong geocoronal emission; this contamination completely swamped the stellar Lyman-$\alpha$ emission and rendered this visit unsuitable for analysis (see Fig. \ref{k2-18_A}). The second visit (B) contained four successful exposures, of which two were performed before the transit ingress and the other two in-transit.

The data were reduced using the standard STIS pipeline, except for the spectral extraction. Since the star is faint, the automated extraction is unable to accurately find the stellar spectrum in the flat-fielded frames. Furthermore, the dark current background of the FUV-MAMA detector of STIS reaches levels high enough to be comparable with the stellar spectrum. In order to correctly extract the spectrum and remove the dark current background, we use the \texttt{x1d} method of \texttt{stistools}\footnote{\footnotesize{Software freely available at \url{https://stistools.readthedocs.io/en/latest/}.}} with user-defined values for the location of: i) the spectrum in the cross-dispersion direction, and ii) the regions where  the dark current background near the spectrum can be accurately estimated. 

\begin{figure}
\centering
\includegraphics[width=\hsize]{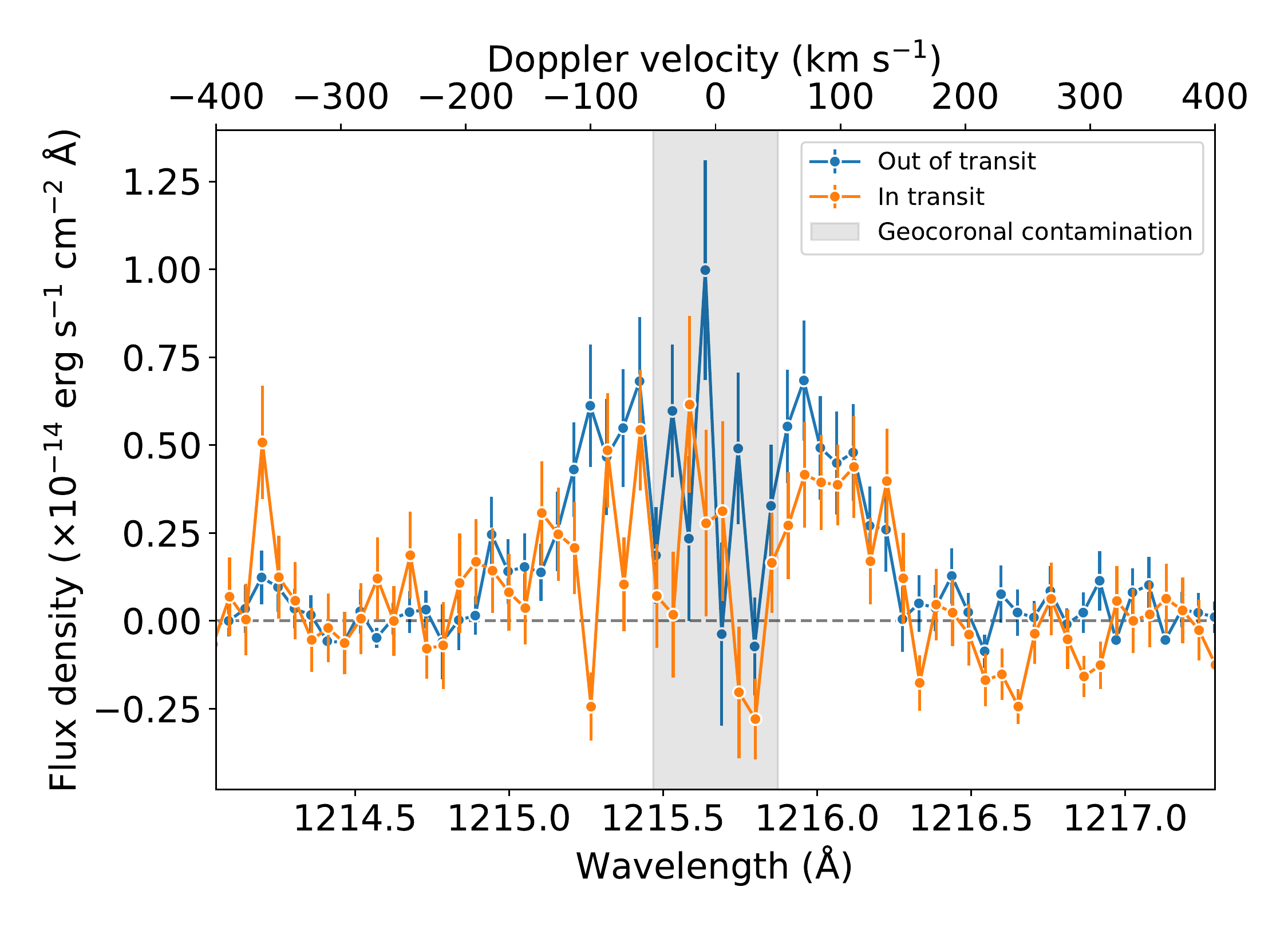}
\caption{\emph{HST}/STIS spectra of K2-18 in Visit B. The Doppler velocities are in the stellar rest frame. The shaded interval is the region with geocoronal contamination.}
\label{k2-18_spec}
\end{figure}

Using visual inspection, we determined the location of the spectrum in the cross-dispersion direction to be $y = 389$ px (parameter \texttt{a2center} in the pipeline). Determining the best location of the background is not as straightforward; normally, the pipeline extracts the background from regions far from the spectrum, but these regions have discrepant levels of dark current compared to the region of the spectrum. Therefore, we chose to use regions immediately near the location of the spectrum to determine the background (parameters \texttt{bk1offst} and \texttt{bk2offst} in the pipeline), namely at a distance of $\Delta = 20$ px from \texttt{a2center}. We combine the two out-of-transit spectra and the two in-transit spectra separately in order to isolate potential signals of an atmospheric signal of K2-18~b, and the resulting spectra are shown in Fig. \ref{k2-18_spec}.

\section{Tentative detection of a H-rich exosphere in K2-18~b}

Since the interstellar medium (ISM) absorbs the core of the Lyman-$\alpha$ emission line, we can only observe the attenuated fluxes in the blue and red wings of the stellar line. We integrated the flux densities in wavelength space between Doppler velocities [-160, -50] km s$^{-1}$ and [+50, +160] km s$^{-1}$, respectively, to produce the Lyman-$\alpha$ light curves in the blue and red wings. 

\begin{figure*}
\centering
\begin{tabular}{cc}
\includegraphics[width=0.44\hsize]{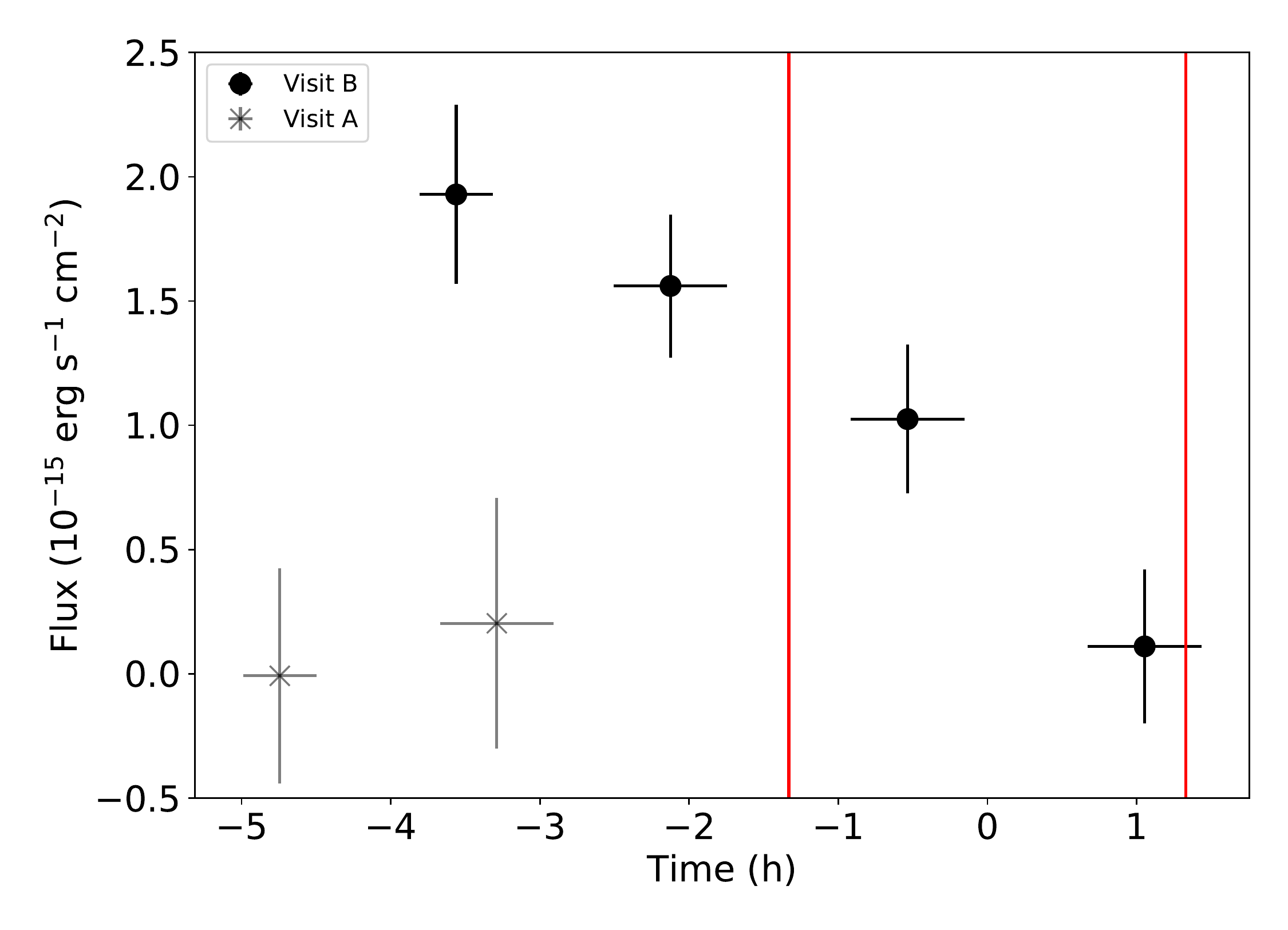} & \includegraphics[width=0.44\hsize]{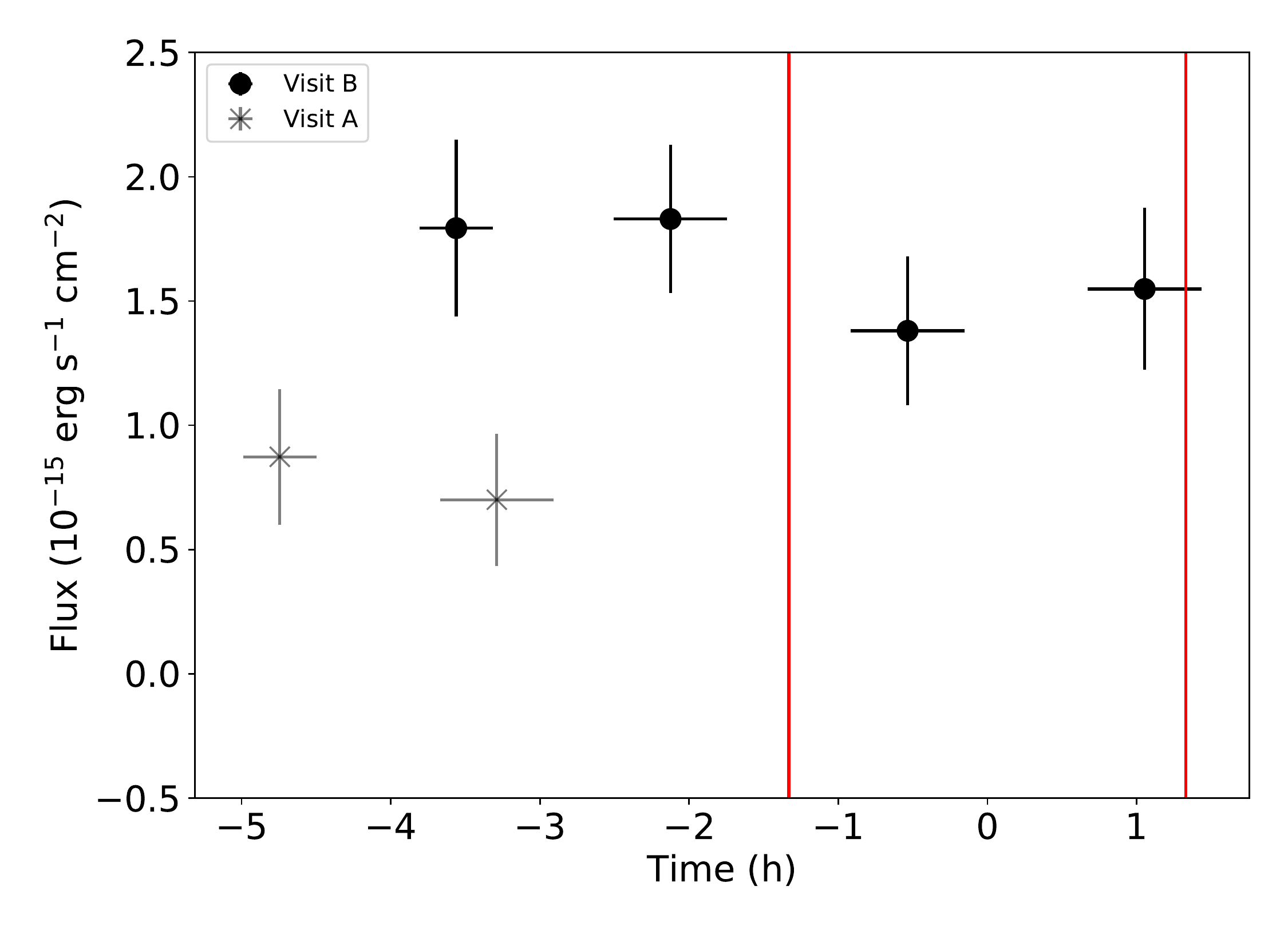}
\end{tabular}
\caption{Light curves of the blue (left panel) and red (right panel) wings of the Lyman-$\alpha$ emission of K2-18 during the transit of planet b. $\mathrm{Time} = 0$ corresponds to the transit center according to the ephemeris of \citet{2019ApJ...887L..14B}. The vertical red lines represent the ingress and egress of K2-18~b. Visit A was affected by strong geocoronal contamination, and therefore the measured stellar fluxes are likely inaccurate.}
\label{k2-18b_lc}
\end{figure*}

During Visit B, we observe a steep decrease in the blueshifted \lya\ fluxes during the transit of K2-18~b (left panel in Fig. \ref{k2-18b_lc}), reaching almost zero emission near the planetary egress. This decrease in flux is also seen when we compare the combined out-of-transit and in-transit spectra (see Fig. \ref{k2-18_spec}). The redshifted \lya\ emission of the combined in-transit orbits varies by $14\% \pm 23\%$, which is consistent with a stable redshifted flux during the visit; this indicates that the variation in the blue wing is likely to be astrophysical in nature. Similar results are obtained even when we extract the background emission at different positions in the detector.

The blue wing flux measured in the combined in-transit spectra decreases by $67\% \pm 18\%$ in relation to the combined out-of-transit spectra; in particular, the last orbit displays an absorption of $93\% \pm 18\%$ in relation to the combined out-of-transit spectra. Although statistically significant, we conservatively deem this result tentative until it is repeated in future observations; for a reference, the intrinsic stellar variability of the \lya\ emission of HD~97658~b is on the order of a few tens of percent at $\sim$2$\sigma$ confidence \citep{2017A&A...597A..26B}. If confirmed to be linked to the transit of K2-18~b, the variation in \lya\ flux can be interpreted as the absorption caused by an extended, H-rich exosphere of the planet. Such a large absorption signal can  be explained by a combination of large atmospheric escape rate and long photoionization lifetime of the H atoms in the exosphere. This result gives further support to the hypothesis that K2-18~b possesses a H$_2$-dominated envelope \citep[as in the conclusions of][]{2019ApJ...887L..14B}, unlike the super-Earths 55~Cnc~e and HD~219134~b. Previous results for the super-Earths HD~97658~b and GJ~1132~b were inconclusive due to stellar variability for the first \citep{2017A&A...597A..26B} and lack of stellar blue wing emission in the second \citep{2019AJ....158...50W}.

\section{The high-energy environment of K2-18~b}

Determining the high-energy environment of K2-18~b provides a critical piece of information to interpret the evolution and current state of its atmosphere. To that end, we used the STIS observations of K2-18 to reconstruct its intrinsic \lya\ spectrum and estimate the high-energy irradiation received by planet b. The reconstruction process follows the standard method used in, for example, \citet{2017AJ....154..121B} and \citet{2018A&A...620A.147B}. In short, we fit the observed spectrum to a model of the intrinsic emission line attenuated by ISM absorption, scaled for distance and convolved with the instrumental response; the fit yields an estimate of the intrinsic emission and certain properties of the ISM in the line of sight. In this process, we assume that the intrinsic \lya\ emission of K2-18 possesses a Gaussian profile \citep[applicable for M dwarfs and for the quality of the available spectra; see][]{2017A&A...602A.106B, 2018A&A...620A.147B}, and fix the temperature and turbulent velocity of the ISM to 8000~K and 1.23~km~s$^{-1}$ \citep[for the NGP cloud, as estimated by the LISM calculator][]{2008ApJ...673..283R}, respectively. We also set the deuterium-to-hydrogen ratio ($\ion{D}{I}/\ion{H}{I}$) to $1.5 \times 10^{-5}$ and the systemic velocity to 0.6537~km~s$^{-1}$ \citep[measured with high-resolution spectra by][]{2017A&A...608A..35C}.

\begin{figure}[h]
\centering
\includegraphics[width=\hsize]{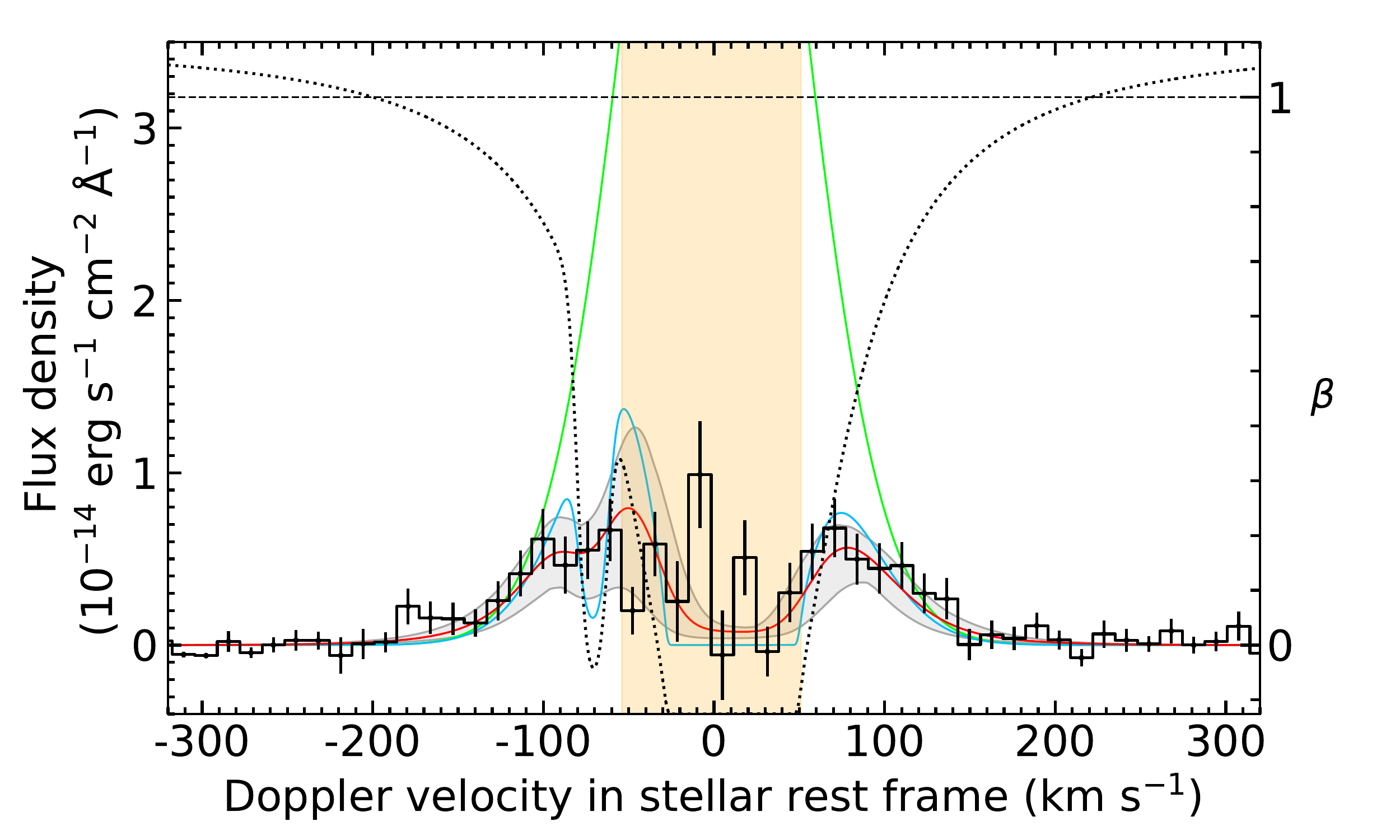}
\caption{Reconstructed intrinsic \lya\ spectrum of K2-18 (green) and the average \lya\ profile as observed with \emph{HST}/STIS (black bars). The red (blue) curve shows the inferred observable spectrum with (without) instrumental convolution and the shaded region represents the 1$\sigma$ uncertainty for the red curve. The dotted curve shows the inferred ISM absorption profile.}
\label{k2-18_rec_spec}
\end{figure}

The result of the \lya\ line reconstruction is shown in Fig. \ref{k2-18_rec_spec}. We simultaneously fit each exposure to a global \lya\ line model within $\pm 300$~km~s$^{-1}$ in the stellar rest frame, excluding the band that corresponds to significant airglow contamination (approximately the range $\pm 50$~km~s$^{-1}$, shown in orange in Fig. \ref{k2-18_rec_spec}). When fitting the in-transit orbits we excluded the pixels corresponding to the range potentially absorbed by the planet. We explore the parameter space using the Markov chain Monte Carlo ensemble sampler implementation of \texttt{emcee} \citep{2013PASP..125..306F}; in addition, we report the uncertainties based on the highest density interval (HDI), which contains 68.3\% of the distribution mass such that no point outside the interval has a higher density than any point within it. We fit for four free parameters in total: the temperature (assuming a Gaussian thermal broadening) and amplitude of the intrinsic stellar \lya\ line, and the radial velocity and \ion{H}{I} density of the ISM. We estimated the EUV flux in K2-18~b using the relation from \citet{2014ApJ...780...61L}.

We determined that the heliocentric radial velocity of the ISM is $V_\mathrm{R} = 11.39^{+4.04}_{-4.73}$ km s$^{-1}$, and the \ion{H}{I} column density in the line of sight is $\log_{10}{\eta} = 18.16^{+0.44}_{-0.34}$ cm$^{-2}$. While $\eta$ is consistent with the results from \citet{2005ApJS..159..118W}, the value we determined for $V_\mathrm{R}$ differs by $2\sigma$ with the value predicted by LISM calculator \citep{2008ApJ...673..283R}.

We estimated the properties of the high-energy environment of K2-18~b based on the reconstructed \lya\ emission and the results are shown in Table \ref{env_prop}; the large uncertainties in these estimates are due to the low signal-to-noise ratio (S/N) of the spectra and the uncertainties in the semi-empirical relations used to estimate the EUV flux. The best fit model of the intrinsic \lya\ emission of K2-18 results in a $\beta$ (ratio between radiation pressure and stellar gravity; see right axis of Fig. \ref{k2-18_rec_spec}) of about 2.2. Although more data would be necessary to confirm this line shape, and the corresponding high $\beta$, it nonetheless suggests that radiation pressure could blow away the escaping hydrogen atoms more strongly in K2-18~b than in GJ~436~b \citep{2016A&A...591A.121B} or the TRAPPIST-1 planets \citep{2017AJ....154..121B}. This value of $\beta$ is similar to that inferred for GJ~3470~b \citep{2018A&A...620A.147B}, which could indicate that the exosphere of K2-18~b possesses a similar shape. Using the energy-limited escape \citep[e.g.,][]{2015A&A...576A..42S} as an initial estimate, we predict that the total escape rate in K2-18~b is two orders of magnitude lower when compared to the values inferred for GJ~436~b \citep[$\sim$2.2 $\times 10^{10}$ g s$^{-1}$;][]{2016A&A...591A.121B} and GJ~3470~b \citep[$\sim$8.5 $\times 10^{10}$ g s$^{-1}$;][]{2018A&A...620A.147B}. We estimated the photoionization rate and lifetime as in \citet{2017A&A...597A..26B}, which yields a value likely above 1200 h for the latter; this long lifetime means that the H atoms in the exosphere of K2-18~b could stay neutral for much longer than in GJ~436 (12 h) and GJ~3470~b (3.5 h). These results are not surprising because K2-18~b is subject to lower irradiation levels than the aforementioned warm Neptunes. 

The inferred escape rate of K2-18~b is likely underestimated because we do not take into account the stellar X-ray flux, which is currently unknown. For a similar star like GJ~436, the ratio between X-ray and EUV emission is $\sim$0.23 \citep{2016A&A...591A.121B}, which provides an upper limit on the level of underestimation contained in our calculation; however, the contribution of X-ray flux may be less important for the ionization conditions because of the smaller ionization cross-section when compared to EUV wavelengths.

\begin{table}
\caption{Properties of the high-energy environment of K2-18~b.}
\label{env_prop}
\centering
\begin{tabular}{ll}
\hline
\lya\ flux (erg s$^{-1}$ cm$^{-2}$) & $100.7^{+96.1}_{-82.4}$ \\
EUV 10-91.2 nm flux (erg s$^{-1}$ cm$^{-2}$) & $107.9^{+124.7}_{-90.8}$ \\
Photoionization rate ($\times 10^{-6}$ s$^{-1}$) & $3.7^{+7.5}_{-3.5}$ \\
Photoionization lifetime (h) & $3600^{+49900}_{-2400}$ \\
Escape rate at 100\% efficiency ($\times 10^8$ g s$^{-1}$) & $3.5^{+4.0}_{-2.9}$ \\
\hline
\end{tabular}
\end{table}

\section{Conclusions}

K2-18~b is currently one of the best targets for transit spectroscopy among sub-Neptune planets due to its large scale height, its short distance from the Sun, and the infrared brightness of the host star. Previous results have shown evidence that the atmosphere of the planet is dominated by H$_2$/He and contains water vapor. In these atmospheric conditions and under the expected high levels of EUV irradiation, K2-18~b is prone to efficiently losing its atmosphere and producing a detectable excess absorption of H in \lya\ caused by a H-rich exosphere during transit. In this study we analyzed four \emph{HST} orbits before and during the transit of K2-18~b with the STIS instrument to search for this feature.

We analyzed the flux time series of both the blue- and redshifted wings of the stellar \lya\ emission. The blue wing displays a significant excess absorption during the transit; in particular, near the egress of K2-18~b, the flux in the blue wing is consistent with 100\% absorption. The in-transit red wing fluxes vary by $14\% \pm 23\%$ and are significantly more stable than the blue wing fluxes. A blueshifted absorption could indicate the presence of a H-rich exosphere around K2-18~b being swept away by radiation pressure from its host star towards the direction of the observer, similar to the exospheres of GJ~436~b and GJ~3470~b.

Despite the low S/N of the observed spectra, we were able to reconstruct the intrinsic stellar emission (without the ISM absorption) to assess the high-energy environment of K2-18~b. Our first estimate for the expected total escape rate of K2-18~b leads to a value on the order of $10^8$ g s$^{-1}$. The ratio between radiation pressure and gravity ($\beta$) suggests that the exosphere of K2-18~b is in a similar state to that observed for GJ~3470~b. We estimate that the EUV ($10-91.2$ nm) flux in the planet is on the order of $10^1-10^2$ erg s$^{-1}$ cm$^{-2}$. At the estimated escape rate, it is likely that the planet will lose only a small fraction (1\% or less) of its mass during its remaining lifetime, and therefore it is probably not an archetypal planet crossing the radius valley to become a bare rock  \citep{2017AJ....154..109F, 2018MNRAS.479.4786V, 2018AJ....156..264F}; as such, the planet will likely retain its volatile-rich atmosphere due to the more amenable EUV irradiation flux than for example GJ~3470~b, which is at least ten times more EUV irradiated than K2-18~b.

Since we observed only one partial transit of K2-18~b, we conclude that the H-rich exosphere detection is only tentative for now, and more observations are needed to rule out stellar activity effects and confirm the reported feature. Furthermore, additional observations of the \lya\ spectrum of K2-18 will help in better constraining the high-energy environment of the planet and its atmospheric escape history.

\begin{acknowledgements}
LAdS thanks M. Turbet and M. L\'opez-Morales for the insightful discussions about K2-18~b. This project has received funding from the European Research Council (ERC) under the European Union’s Horizon 2020 research and innovation programme (project {\sc Four Aces}; grant agreement No 724427), and it has been carried out in the frame of the National Centre for Competence in Research PlanetS supported by the Swiss National Science Foundation (SNSF). This research is based on observations made with the NASA/ESA Hubble Space Telescope. NA-D acknowledges the support of FONDECYT project 3180063. The authors are grateful to the anonymous referee for the quick and helpful review. The data are openly available in the Mikulski Archive for Space Telescopes (MAST), which is maintained by the Space Telescope Science Institute (STScI). STScI is operated by the Association of Universities for Research in Astronomy, Inc. under NASA contract NAS 5-26555. This research made use of the NASA Exoplanet Archive, which is operated by the California Institute of Technology, under contract with the National Aeronautics and Space Administration under the Exoplanet Exploration Program. We used the open source software SciPy \citep{scipy_ref}, Jupyter \citep{Kluyver:2016aa}, Astropy \citep{2013A&A...558A..33A}, Matplotlib \citep{Hunter:2007} and \texttt{emcee} \citep{2013PASP..125..306F}.
\end{acknowledgements}

\bibliographystyle{aa}
\bibliography{biblio.bib}

\end{document}